# ABSTRACT

# EFFECTIVENESS OF SESAME OIL FOR THE PREVENTION OF PRESSURE ULCER IN PATIENTS WITH BED REST UNDERGOING HOSPITALIZATION

**Chrisyen Damanik[1], Sumiati Sinaga[2], Kiki Hardiansyah[3]**
[1,2,3]School of nursing, Institute of Health Technology And Science Wiyata Husada Samarinda, Indonesia
Email: chrisyendamanik@itkeswhs.ac.id

Pressure Ulcer is one of the most problems in patients with bed rest. Reposition and skin care are deterrent against the incidence of pressure ulcer. **Objective**: This studi aimed to analize the effectiveness of sesame oil for the prevention of pressure ulcer in patients with bed rest undergoing hospitaization. **Method**: This study used a randomized controlled trial design. Forty samples were devided groups: control and intervention groups. This study was analysed using Chi Square. **Results**: The results showed that there was a significant difference between two group (p=0,04). **Conclusions**: Skin care with sesame oil can prevention of pressure ulcers. These results recommended that sesame oil can be used for nursing intervention for the prevention of pressure ulcers.
Keyword: Pressure Ulcer, Sesame Oil, Bed Rest, Hospitalization

**Introduction**

Pressure ulcers (PUs) are caused by tissue damage when the blood supply to an area of skin is diminished as a result of pressure. Although most pressure ulcers are preventable, all patients are at risk (1). Risk assessment should be carried out as soon as possible and within a maximum of 8 hours of the patient being either admitted to hospital or onto a community caseload; this should be repeated as often as required based on patient acuity (National Pressure Ulcer Advisory Panel (NPUAP).

PUs have become a worldwide concern for health professionals, with the cost burden of managing them and associated complications in excess of £2.1 billion annually (2). Clinical interventions for PU prevention include holistic assessment, risk assessments and preventive measures (3). Nurses have a central role in prevention and management of pressure areas. They should be able to assess patients' risk of developing PUs using evidence-based practice, recognised risk assessment tools and by completing a holistic assessment. Nurses must be able to identify the risk factors associated with developing PUs and implement appropriate measures to deliver harm-free care. Specifically, nurses can play the most important role in assessing PUs risk factors: when patients are admitted into long-term care facilities, nurses observe the injury-prone area in order to identify the preventive early stage and then implement preventive nursing care (4)(5).

Pressure ulcers (PUs) prevention remains a significant challenge for nurses(6). Patients and families know that pressure ulcers are painful and slow to heal. Some risk factors for the development of pressure ulcers/injuries include advanced age, immobility, incontinence, inadequate nutrition and hydration, neuro-sensory deficiency, device-related skin pressure, multiple comorbidities and circulatory abnormalities (7)

The incidence of pressure ulcers in adults varies from 0 to 12% in acute care settings, 24.3 to 53.4% in critical care settings and 1.9 to 59% in elderly care settings (8). When caregivers practice the best care every time, patients can avoid needless suffering. Pressure area care is an essential component of nursing practice, with all patients potentially at risk of developing a pressure ulcer (9) It is nurses' primary responsibility for maintaining skin integrity (10) and prevention of its complications Recognizing patients at risk of developing PU in early time is an essential part of the prevention care pathway (11).

Sesame oil (sesame oil) is one of the herbs that have effectiveness as an antioxidant, anti-inflammatory and analgesic. Owned analgesic properties due to the content of lignan found in sesame oil, which is able to inhibit the pain-causing chemical mediators such as prostaglandins (12). Sesame oil is processed from sesame seeds which are very rich in protein, vitamins and minerals. In addition, it has nutrients that contain essential fatty acid compounds, omega 6, omega 9, antioxidants, which function to regulate the balance of the immune system, inhibit the inflammatory process (13). Referring to this phenomenon, the research objective was to determine effectiveness of sesame oil for the prevention of pressure ulcer in patients with bed rest undergoing hospitaization.

**Method**

This study is a randomized controlled clinical trials (RCT). The design used is a parallel design without matching. The location of this research is Aji Batara Dewa Sakti Samboja. This Research is conducted in July to August 2020. The sample is selected by consecutive sampling, the number of samples involved in the study was 40 people, consisting of 20 people the intervention group and 20 control group, with the inclusion criteria**:** 1) Inpatients who experience bed rest and willing to become respondents, 2) The risk of pressure ulcer is assessed using the NPUAP scale, 3) Using the standard bed and mattress used in the treatment room 4) Shows negative results for an allergy test and, 5) Not undergoing special treatment for pressure ulcer. The allocation of samples into the intervention group and the control group performed the randomization techniques.

Data collection tool was a questionnaire containing questions related to the characteristics of the respondent, which contains the characteristics of the respondent containing questions including: age, gender, body mass index, albumin levels, smoking status as well as an observation sheet on the pressure ulcers measure scale using the Press Wound Stadium based on NPUAP 2009.

In this study, the researcher was assisted by a research assistant who was a nurse who served at the Samboja hospital and had passed the interrater reliability test which guaranteed the similarity in the perception of the observations between the researcher and the research assistant. As for the technical procedures in the implementation of the research, the following actions were taken: Each sample selected by the researcher, both the control and intervention groups, would be conditioned to receive standard treatment measures in the prevention of pressure sores, namely changing positions every 2 hours, namely tilted left-right and supine. When tilted left and right, the patient is supported by 1 pillow on the head, shoulders, and between the knees so that the ischium and sacrum are lifted $30^0$, and bathed twice a day morning and evening using a washcloth and soap. For the intervention group, the researchers added a light massage (backrub) treatment using sesame oil on the back from the scapula (shoulder) to the ischium and the hell (heel) to Malleolus area 2 times after bathing. The duration of this research and observation is 3 days for each sample person. The basis for the consideration of this study was carried out for 3 days referring to previous studies and referring to the average length of stay of patients who experience bed rest at Samboja Hospital around 5-8 days of treatment.

Analysis of the data in this study include univariate and bivariate. Univariate analysis describes the characteristics of each of the variables studied. Presentation of each variable by using tables and interpretations based on the results obtained. Bivariate analysis were conducted to prove the hypothesis The statistical test used for bivariate analysis was Chi Square

**Results**
**a. Univariat Analysis**
**Table 1**. Distribution of respondents based on the characteristics of the respondents at the Aji Batara Agung Dewa Sakti hospital, Samboja. July-August 2020 (n1=n2=20)

| Variable | Categori | Intervention group (n=20) | | Control Group (n=20) | |
|---|---|---|---|---|---|
| | | f | % | f | % |
| Age | < 50 year | 10 | 50 | 12 | 60 |
| | ≥ 50 year | 10 | 50 | 8 | 40 |
| Gender | Male | 7 | 35 | 8 | 40 |
| | Female | 13 | 65 | 12 | 60 |
| Smoking History | No Smoke | 13 | 65 | 13 | 65 |
| | Smoke | 7 | 35 | 7 | 35 |
| Albumin Levels | < 3gr/dl | 8 | 40 | 6 | 30 |
| | ≥ 3 gr/dl | 12 | 60 | 14 | 70 |
| Body Mass Index | < 18 kg/m$^2$ | 10 | 50 | 5 | 25 |
| | ≥ 18 kg/m$^2$ | 10 | 50 | 15 | 75 |

**Table 2.** Distribution of Respondents Based on the Incidence of Pressure Ulcers at Aji Batara Dewa Sakti Samboja Hospital, July-August 2020 (n1=n2=20)

| Stage of Pressure Ulcers | Intervention (n=20) | | | | Control (n=20) | | | |
|---|---|---|---|---|---|---|---|---|
| | Pre Test | | Post Test | | Pre Test | | Post Test | |
| | f | % | f | % | f | % | f | % |
| 1. No Pressure Ulcers | 0 | 0 | 16 | 80 | 0 | 0 | 6 | 55 |
| 2. Stage I | 10 | 50 | 4 | 20 | 10 | 50 | 9 | 30 |
| 3. Stage 2 | 10 | 50 | 0 | 0 | 10 | 40 | 5 | 15 |

shows that most of the respondents in the intervention group (80%) experienced a decrease in the degree of pressure sores after the intervention with sesame oil was administered. The results of the homogeneity test for the variable characteristics of the respondents, age, sex, smoking history, albumin levels and body mass index, had an equivalent (homogeneous) variance.

**b. Hasil Analsis Bivariat**
Table 3. The difference in the incidence of pressure ulcers between the intervention group and the control group in the Aji Batara Dewa Sakti Samboja Hospital Care Unit (n1=n2=20)

| Stage of Pressure Ulcers after Intervention | Group | | Total | OR CI | p-value |
|---|---|---|---|---|---|
| | Intervention | Control | | | |
| No Pressure Ulcers | 16 (11) | 6 (11) | 22 | 9.333 (2.180-39.962) | 0,04 |
| Stage 1+2 | 4 (9) | 14 (9) | 18 | | |

**Discussion**
There is a difference in the incidence of pressure ulcers after the intervention between the intervention group and the control group. The statistical test results obtained p value = 0.04 (p <α = 0.05), it can be concluded that there is a difference in the proportion of the incidence of pressure ulcers between respondents who were given preventive treatment using sesame oil and without using sesame oil. From the results of the analysis, the OR value of 9.333 means that respondents who are not given sesame oil intervention with a combination of 300 oblique beds will have a 9.333 times chance for the incidence of pressure ulcers compared to respondents who are only given a 300 oblique lying position.

In this study, the herbal therapy used was the use of sesame oil. Sesame oil is processed from sesame seeds which are very rich in protein, vitamins and minerals. In addition, it has nutrients that contain essential fatty acid compounds, omega 6, omega 9, antioxidants, which function to regulate the balance of the immune system, inhibit the inflammatory process (13) Sesame oil also contains a number of lignans: sesamin, epicesamine and sesamolin. Another research result that explains the effectiveness of sesame oil was found by Hirsch et al. (2008) (14), in a study comparing the effectiveness of sesame oil and flamazine ointment in treating superficial burns. Forty respondents were involved in this study, which consisted of two groups. Each group was observed for pain, inflammation and repair of the skin layer. The findings of this study indicate a significant difference between the two groups. In the intervention group that used sesame oil ointment (sesame oil) was significantly effective in preventing pressure sores and reducing the degree of pressure sores, inflammation and repairing the skin layer than the control group.

**Conclusion**

Skin care with sesame oil can prevention of pressure ulcers. These results recommended that sesame oil can be used for nursing intervention for the prevention of pressure ulcers.

**Acknowlegements**

Acknowledgments to the Ministry of Research, Technology and Higher Education, Directorate General of Research and Development Strengthening for providing research grants for funding in 2020. As well as Aji Batara Dewa Sakti Hospital which has granted research permission during the Covid-19 Pandemic